\documentclass[12pt]{article}
\usepackage{times}
\begin{document}
\pagestyle{plain}
\hsize = 6.5 in 				
\vsize = 8.5 in		   
\hoffset = -0.5 in
\voffset = -0.5 in
\baselineskip = 0.32 in	

\def\vF{{\bf F}}
\def\vJ{{\bf J}}

\title{Small Open Chemical Systems Theory: Its Implications
to Darwinian Evolution Dynamics, Complex Self-Organization 
and Beyond\footnote{This article is based on a position paper
submitted to ISTAR-NSF-NSA Workshop on ``Mathematical Foundations of Open Systems'' held at University of Pennsylvania from 
May 23rd--25th, 2010. See http://istarpenn.org/events/pastevents.html}}

\author{Hong Qian\\[13pt]
Department of Applied Mathematics, University of Washington\\
Seattle, WA 98195-3925, USA\\
and\\
Kavli Institute for Theoretical Physics China (KITPC)\\
at the Chinese Academy of Sciences\\
Beijing 100080, PRC
}

\maketitle

\begin{abstract}
The study of biological cells in terms of mesoscopic, nonequilibrium, 
nonlinear, stochastic dynamics of open chemical systems provides 
a paradigm for other complex, self-organizing systems 
with ultra-fast stochastic fluctuations, short-time deterministic nonlinear 
dynamics, and long-time evolutionary behavior with exponentially
distributed rare events, discrete jumps among punctuated equilibria, 
and catastrophe.
\end{abstract}

\section{Introduction}

Post-genomic biology can only be fully understood from a combined
cellular-molecular and evolutionary perspective.   It is an integration of
extremely small and extremely large scale dynamics in space and time
with significant heterogeneity in the system.  These are challenges any 
researcher concerned with small open systems has to confront:  A system
being open and nonequilibrium means its environmental influences at a large scale, usually through a boundary, are felt  to a significant degree by many 
of the small components inside the system. On the other hand, the ultra-fast
dynamics on the smallest scale, with large degrees of freedom, 
appear to be ``stochastic''.  The term
``mesoscopic'' is used in this context in the present work \cite{aqtw,zqq}.
Interestingly, the fundamental issues in both cellular and evolutionary 
dynamics are now firmly cast in terms of a common 
stochastic mathematics \cite{gqq,aoping,wangjin}. 
While the field of stochastic dynamics has more than 100 years of history,
starting with Einstein's 1905 study of Brownian motion and Langevin's
1908 proposal of a stochastic differential equation, we still know
little about it beyond the existence proofs and formal constructions.
In particular, the interplay between nonlinearity and stochasticity giving
rise to emergent phenomena  is still not fully understood \cite{qian_qb}.

Applied stochastic dynamics, as a mathematical tool, is in a rather 
primitive state compared to our deeper understanding of the 
deterministic dynamics that is at the foundation of 300
years of the Newton-Laplace 
world view.  An applied stochastic dynamic theory with in-depth 
understanding of the interaction between deterministic (i.e., drift)
and stochastic (i.e., random) elements is urgently needed for
the progress of biology, and for the understanding of other complex,
open systems. In other words, we are particularly interested in how
stochastic random motion is ``coupled'' to deterministic, often nonlinear, dynamics.  When one observes a noisy oscillation, is it a limit cycle 
with some noise or a rotational random walk? 
These questions lead us to study the concept
of entropy in stochastic dynamics.  It can be shown that the entire
statistical thermodynamic structure, with its core concepts like 
entropy, free energy, and the Second Law, is in fact 
a mathematical one --- molecular
thermal physics turns out to be one example. One can equally develop a
``thermo''-dynamic theory for stochastic cellular processes, for
evolutionary dynamics, and for the dynamics of other mesoscopic 
open systems.

	In this article, I shall first give a brief introduction to
the current status of the mesoscopic open-chemical system theory.  
Then I shall discuss a few topics which, I believe, deserve further 
development.  For more authoritative reviews of the vast literature,
see \cite{jarzynski,seifert,qian_arbp,chowd,goutsias}.

\section{What is the Open Chemical Systems Theory?}

When one leaves an aqueous chemical reaction system alone, one
observes it goes through a transient process
with the concentrations of the chemicals changing with time. This is
known as relaxation kinetics.  Eventually,
it settles into a chemical equilibrium.  If the system is sufficiently
small, then one can also observe the concentrations continuously 
fluctuate with time, though all their statistics are time invariant 
\cite{elson,weissman}.  
Chemists call this state a chemical {\em equilibrium} steady state.

	A living cell as a biochemical system, to
a first order approximation, is in a steady state, known as
homeostasis \cite{homeo}.  It is not in an  equilibrium, however: 
one has to continuously provide a ``nutritious medium'' 
in which the cell ``lives''.
The medium has to contain ``high calorie'' chemicals as
food for the cell, which in turn returns its ``waste product'' to
the medium.   So how should one rigorously define such a chemical
state of a system which is {\em open to exchange chemical
energy and materials with its surroundings} \cite{bertalanffy}?   
This is the motivation for developing the open chemical system 
theory \cite{qian_bpc_05,qian_jpc_06,qian_arpc_07}.  
Such a state of a system is called a {\em nonequilibrium steady 
state} (NESS); it is an example of {\em self-organizing dissipative 
structure} \cite{NP_book}.  With the stochastic dynamic
perspective, one immediately
realizes that a deterministic periodic chemical oscillation
is also an NESS \cite{qian_pnas_02}.

	Driven by the environment, chemical fluxes are 
continuously going through such 
systems.  And because of the presence of nonlinear chemical
reactions, these fluxes are coupled to internal (bio)chemical 
reactions.  Therefore, according to Kirchhoff's law, there 
will be a  myriad of cycle fluxes in an NESS \cite{hill_book}.   
This is why a living cell contains all kinds of metabolic cycles.
A deep mathematical theorem shows that, in a large class of 
stochastic systems, a cycle flux exists if and only if the
rate of entropy production is positive \cite{jqq_book,ge_adv_math_14}.  
When some of the cycle fluxes are sufficiently 
strong, a macroscopic oscillation emerges \cite{gqq_mbs_08}.  
  	
	The existence of the cycle flux, or circulation, turns out to be 
a fundamental property of open chemical systems.  It is 
intimately related to the breakdown of {\em detailed balance}.   
It is mathematically related to a non-self-adjoint
generator for Markov processes.  The presence of cycle flux also indicates
that the stationary probability distribution in one part 
of the system can be influenced by another 
part far away.  Thus the dynamics are nonlocal.  This is not
possible for a system with detailed balance, which 
yields Boltzmann's Law.  Even more important:
a strong flux is associated with a deterministic
kind of motion (breaking a symmetry).  Hence, a 
severely driven chemical system 
can exhibit rhythmic dynamics.  Chemical energy can 
suppress fluctuations in small open chemical systems, making
them behave like little machines \cite{seifert,vonhippel,alberts}.

	In the past several years, people have shown that
almost every biochemically important cellular
function is intimately related to the open chemical
system setting, and correlates with the energy expenditure 
associated with the system.  This includes kinetic proofreading
\cite{hopfield_pnas_74,ninio_75,qian_jmb_06,leibler}, cellular
signal switching \cite{qian_cooper_08}, fidelity
in DNA replication \cite{cady_qian_09}, chemotactic
adaptations, and other biochemical computations \cite{tuyh,mehta,jiacheng}.  
Another significant insight is the
emergence of chemical multi-stability. In stochastic dynamic 
terms, this is characterized by a stationary probability 
density
function exhibiting multiple peaks. We say there are 
multiple stochastic attractors.   These attractors are
emergent properties of dynamics of an open chemical system.
Often, though not always, they are the stochastic 
counterparts of the deterministic stable
attractors \cite{kauffman_book}.  
These attractors define an emergent, 
discrete multi-state
stochastic dynamics on an entirely different time
scale \cite{aqtw}; a {\em cellular evolution time scale} emerges
\cite{ge_qian_jrsi}. 
Therefore, the open chemical system
model is able to conceptually bridge the
detailed biochemical reactions networks and the emergent
cellular dynamics that reflect differentiation,
apoptosis, and epigenetic switching \cite{qian_jcst_10}!
Furthermore, it can be shown that there is also
an emergent ``landscape'' \cite{wolynes}, akin to the adaptive
landscape in evolution theory.  This landscape
is not locally determined; it is  itself an emergent property, and
its predictive power is, in some sense, 
only {\em retrospective} \cite{qian_ge_mcb}.

\section{Nonequilibrium Steady-State,
 Emergent Landscape and Flux in Open Systems}

The concept of a {\em nonequilibrium steady state}, 
also known as a {\em nonequilibrium stationary state},
is one of the central concepts in studies of open chemical systems.
It deserves a focused and in-depth investigation.  Note
that for a large class of stochastic dynamics, there is a 
unique, asymptotically attractive stationary process 
(invariant measure). That is, the system is always self-organizing.  
In statistical physics and the
mathematical theory of interacting particle systems,
related research has been pursued for many years.
The emphasis has been
on the system's infinite size limit while assuming the individuals
are significant ``homogeneous''.   Stochastic processes with NESS
have not been widely studied.  The technical difficulties are mainly in the 
non-symmetric nature of the process, which makes
even the mathematical existence proof impossible.  The 
recent exciting development of the {\em fluctuation
theorem} is precisely along this line \cite{seifert,searles_arpc_08,jarzynski}.  
It is an important first step.  Still, this result is largely unknown in the field of 
stochastic processes \cite{ge_jiang,ft_cmp_08,wookim}. 

	Using a stochastic differential equation 
with small noise ($\epsilon$) as a model system, it will
be important
to further explore the connection between the 
NESS invariant density $f^{ss}_{\epsilon}(x)$, 
its $\epsilon$-dependence, 
and its Lyapunov stability in the corresponding deterministic dynamics. 
Note that while the support of  $f^{ss}_{\epsilon}(x)$ is 
highly singular, with Dirac measure in the limit of $\epsilon$ tending
to zero, the quantity
\[
	\lim_{\epsilon\rightarrow 0} \epsilon \log f^{ss}_{\epsilon}(x)
\]  
is often better behaved, with full support in the space \cite{ge_qian_jrsi}.  
This line of research is intimately related to the theory developed by
Freidlin and Wentzell \cite{freidlin_book}.  This theory needs to be 
simplified and delivered to the hands of broader applied mathematicians 
and scientists.  

	One should not forget, however, that the landscape \cite{wolynes} 
and the stationary distribution are only half of the characterization of the 
system.  The NESS is also characterized by its flux field $\vJ$
which is divergence free; that is $\nabla\cdot\vJ=0$.  Our understanding
of this aspect of stochastic dynamics is still missing \cite{wangjin_pnas_08}.  
In fact, a cogent stochastic interpretation of the $\vJ$ is still elusive.
Nevertheless, the $\vJ$ gives the ``dynamical information''
of an open system in its NESS.  A connection between this line of inquiry 
and the theory of Hodge decomposition and 
algebraic topology, can be found in \cite{qian_wang_cmp,kalpazidou_book}.

\section{The Thermodynamic Structure of
Stochastic Dynamical Systems}

	Historically, thermodynamics has been one of the 
most important organizational principles for systems with
large numbers of atoms and molecules | systems considered
to be complex mainly due to their large number of 
components.   When a system has a large number of 
components, especially when all the components are not
homogeneous, the interactions between components
can only be characterized in a statistical sense.  In fact,
only statistical characterization is meaningful.  We today
recognize that thermodynamics is not merely a 
physical theory about atoms and molecules.  Rather,
any stochastic system characterized in terms of 
Markov dynamics possesses a ``thermo''-dynamic
structure.  Since this theory has nothing to do
with the temperature {\em per se}, we put the
``thermo'' in quotation marks \cite{ge_qian}.    
In essence, one has resolved what Gian-Carlo Rota 
considered ``a standing between physicists and 
mathematicians that  thermodynamics cannot be axiomatized''
\cite{rota}.

	There are two inter-related threads in this general theory: 
one centers around relative entropy (or Kullback-Leibler 
divergence), which is 
intimately related to the free energy in Gibbs' statistical
mechanics $F$ \cite{cover_book,mackey_book}.  An important
fact about $F$ is $dF/dt = -f_d(t) \le 0$.  The other thread
centers around Gibbs-Shannon entropy $S$, which satisfies a
conservation law $dS/dt = e_p(t) - h_d(t)$.  An important
fact here is $e_p(t)\ge 0$.   The terms  
$f_d$, $e_p$, and $h_d$ are called free energy dissipation rate, 
entropy production rate, and heat dissipation rate, respectively.

	For Markov systems with detailed balance, which correspond
to closed systems: $f_d(t)=e_p(t)$.  Furthermore, in the long
time limit, $f_d=e_p=h_d=0$.  This is an equilibrium
steady state (with fluctuations).

	For Markov systems without detailed balance, which
correspond to open systems: $e_p(t) = f_d(t)+Q_{hk}(t)$ where
$Q_{hk}(t)\ge 0$.  $Q_{hk}$, called house-keeping heat,
or adiabatic entropy production \cite{ge_qian,esposito},
characterizes the amount of energy expenditure (i.e., a battery)
that sustains the system away from equilibrium.  In the
long time limit, $f_d=0$ but $e_p=Q_{hk}=h_d > 0$.  The
entropy production rate $e_p$ characterizes total irreversibility.
It has two distinctly different origins: the system's spontaneous 
relaxation (organization) $f_d$ and the external environmental 
drive $Q_{hk}$ \cite{ge_qian}.   We have recently suggested
to read the above equation $Q_{hk}(t)-e_p(t)=-f_d(t)=dF/dt$
as a novel balance equation, for free energy $F$ \cite{qian_jmp}.

	This theory has only been presented in an applied
mathematics style.  Further theoretical mathematical analysis
is necessary on the subject, as well as applications using this 
structure to gain further understanding of stochastic dynamics.  
In addition, issues such as the ``principle of maximal entropy
production'' and related topics have never been seriously
investigated with mathematical rigor until recently 
\cite{ge_dill,dill,polettini}, even though they
are actively discussed in applied fields such as climate science and 
ecology \cite{dewar_jpa_05,martyushev_05,paltridge_07}, as
well as in high-profile headlines \cite{whitfield}.

\section{Investigation and Characterization of
Complex Systems and Their Dynamics in terms of
Nonlinear, Stochastic Models}

	Sections 3 and 4 set up the
fundamental mathematics for studying complex, mesoscopic
open-system dynamics.  This approach has shown
great promise and has provided some powerful novel
ideas, as well as a deeper understanding
of small open-chemical systems.    Whether this approach
can be applied to other open systems in biology, economics,
and beyond remains to be investigated.  One 
particular advantage of chemical systems is the existence
of the theory of Delbr\"{u}ck-Gillespie processes, together with its
{\em chemical master equation}.  
This is the stochastic generalization of the Law of 
Mass Action from classical, nonlinear chemical
kinetics.  It is a combination of the stochastic framework above
and this particular version of stochastic chemical dynamics
that has offered us insights on open, fluctuating biochemical 
systems through concrete mathematical models.

	Applying this approach to other systems is
essential to extending our understanding of mesoscopic open 
systems, and all the topics covered in this special issue.   
Several other research carried out under different headings 
also fit in this general scheme.  One of them is the area of {\em stochastic
resonance}  (SR).  Indeed, SR has offered a great deal
to our current understanding of the important issues.   But
one needs to broaden the scope: as we have pointed out,
``oscillations'' in an NESS is a necessity.  They should not 
be considered a peculiar type of behavior \cite{zqq,qian_prl_00}.

\section{Education: Stochastic Dynamics beyond Brownian Motion}

	There is no doubt that 
Brownian motion is one of the most important aspects of
stochastic dynamics in continuous space and time.
It offers a complete departure from Newton-Laplace's
classic view of a ``smooth world'', and has yielded deep insights
in connection with geometry and other branches of high
mathematics, as recognized by the Fields medals in 2006.

Still, by completely focusing on this aspect of stochastic
dynamics, our understanding of a general stochastic dynamical
system which contains both stochastic and nonlinear deterministic
elements is rather rudimentary. 
In particular, most mathematics are built around the 
well-studied {\em symmetric Markov processes}, which
have corresponding self-adjoint linear operators. 

	As we have discussed, open systems, when represented
in terms of Markov processes, are precisely non-symmetric processes. 
This is one of the lessons we learned from the open-chemical systems
theory.  The nonsymmetricity can be quantified
by  {\em entropy production} \cite{jqq_book,ge_adv_math_14}.   
For discrete-state Markov
processes, the symmetric processes are equivalent to 
Kolmogorov's condition \cite{kolmogorov_31}.  Concepts 
such as cycle condition, detailed balance, dissipation and 
irreversible entropy production have all been independently
discovered in chemistry: Wegscheider's relation in 1901
\cite{weg_zpc_01}, detailed balance by G.N. Lewis in 
1925 \cite{lewis_pnas_25}, Onsager's dissipation function
in 1931 \cite{onsager_31}, and the formulation of 
entropy production in the 1940s \cite{prigogine,tolman_rmp}.

	Compared to the college education of deterministic
mathematics, stochastic mathematics is also  embarrassingly behind.
Here is an example:  while most of my third-year students are familiar with
the notion of a random variable $X$ with probability density function
$f_X(x)$, most of them have never learned, nor are able to work out, how to 
compute the distribution of $Y=g(X)$, assuming $g$ is a monotonic
function!    There is no reason for this: stochastic mathematics
is not intrinsically harder if one only deals with discrete events.  In
particular, with the power of computing, one should be able to teach
stochastic thinking to first-year students in science, engineering, 
economics, and  social science.   We need to have accessible educational programs. Here I emphasize stochastic thinking as distinctly
different from statistical thinking.  It is mechanistically motivated 
mathematical deduction rather than data driven. 

	Finally, I would like to state that the stochastic 
dynamical theory is not an alternative to the deterministic one.
It is a more complete description of nature, which is
capable of representing systems with and without uncertainties.

	{\bf Acknowledgements.}  I thank 
Jacob Price and Lowell Thompson for carefully reading the manuscript.

\end{document}